\documentclass[letterpaper]{article} 
\usepackage[draft]{aaai25}  
\usepackage{times}  
\usepackage{helvet}  
\usepackage{courier}  
\usepackage[hyphens]{url}  
\usepackage{graphicx} 
\usepackage{natbib}  
\usepackage{caption} 
\usepackage{booktabs}
\usepackage{graphicx}  
\usepackage{arydshln}
\usepackage{algorithm}
\usepackage{algorithmic}
\usepackage{subcaption}
\usepackage{amsmath}
\usepackage[dvipsnames]{xcolor}
\usepackage{hyperref}
\hypersetup{
colorlinks=true,
linkcolor=red,
citecolor=NavyBlue,
urlcolor=magenta,
}

\usepackage{newfloat}
\usepackage{listings}
\DeclareCaptionStyle{ruled}{labelfont=normalfont,labelsep=colon,strut=off} 
\floatstyle{ruled}
\newfloat{listing}{tb}{lst}{}
\floatname{listing}{Listing}
\title{Anomaly Detection for Hybrid Butterfly Subspecies via Probability Filtering}
\author{
    Bo-Kai Ruan,
    Yi-Zeng Fang,
    Hong-Han Shuai,
    Juinn-Dar Huang
}
\affiliations{
    National Yang Ming Chiao Tung University\\
    {\{\tt bkruan.ee11, joycefang1213.ee11\}@nycu.edu.tw}
}

\begin{document}

\maketitle

\begin{abstract}
Detecting butterfly hybrids requires knowledge of the parent subspecies, and the process can be tedious when encountering a new subspecies. This study focuses on a specific scenario where a model trained to recognize hybrid species A can generalize to species B when B biologically mimics A. Since species A and B share similar patterns, we leverage BioCLIP as our feature extractor to capture features based on their taxonomy. Consequently, the algorithm designed for species A can be transferred to B, as their hybrid and non-hybrid patterns exhibit similar relationships. To determine whether a butterfly is a hybrid, we adopt proposed probability filtering and color jittering to augment and simulate the mimicry. With these approaches, we achieve second place in the official development phase. Our code is publicly available at \url{https://github.com/Justin900429/NSF-HDR-Challenge}.
\end{abstract}


\section{Introduction}

Butterfly species can diverge into different subspecies due to geographical or habitat separation, leading to distinct variations in their visual appearances, such as wing color patterns. Typically, individuals of the same subspecies mate and produce offspring with predictable appearances. However, when subspecies encounter one another in overlapping regions, interbreeding may occur, resulting in hybrid offspring. These hybrids inherit characteristics from both parent subspecies, displaying intermediate visual patterns.

\vspace{5pt}

\noindent This study focuses on developing an anomaly detection algorithm to differentiate between hybrids and non-hybrids within a species. Furthermore, we explore whether such an algorithm can generalize to another species that visually mimics the first, providing insights into the generalization of the provided approach. The mimics represent two distinct species that evolve similar visual patterns when they share geographic ranges. This mimicry can serve as a defense mechanism against predators, particularly when one or both species are toxic or unpalatable.


\section{Methods}

\paragraph{Problem Formulation.} Given an input image of butterfly wings, assign an anomaly score in the range $[0, 1]$. A higher score indicates a higher probability of a hybrid butterfly.

\subsection{Backbone}

Since the challenge requires transferring results from species A to the unseen species B, our model must incorporate a feature extractor capable of distinguishing species while capturing patterns that remain consistent across them. To achieve this, we adopt BioCLIP~\cite{stevens2024bioclip} as our backbone feature extractor, as it recognizes inputs and generates features based on biological taxonomy.

\vspace{5pt}

\noindent To further refine feature extraction for butterfly species, we fine-tune the backbone without freezing its weights. However, updating the backbone parameters may compromise its original generalizability. To preserve its original ability, we apply a smaller learning rate to the backbone, ensuring that updates remain subtle and do not deviate significantly from the original weights.

\begin{figure}
    \centering
    \includegraphics[width=\linewidth]{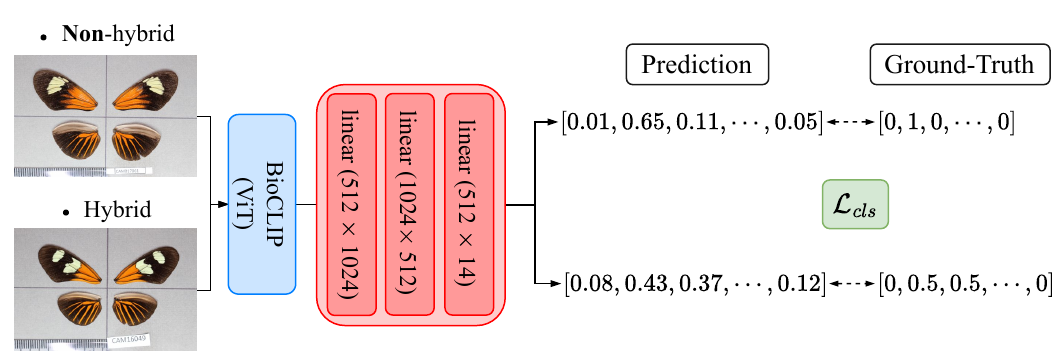}
    \caption{Illustration of the training pipeline.}
    \label{fig:model}
\end{figure}

\subsection{Probability Filtering}

Since the subspecies within a particular species are fixed, each image can be directly classified into a specific subspecies. To achieve this, we incorporate a three-layer classifier into the BioCLIP visual model. For non-hybrid butterflies, we assign a ground-truth class probability of 1 to their respective subspecies. In contrast, since hybrid butterflies originate from two different parent subspecies, we distribute their probabilities equally between both classes, assigning each a value of 0.5. The model architecture is illustrated in Figure~\ref{fig:model}. We train the entire model using cross-entropy loss. After training, we expect the model to predict non-hybrid butterflies with high confidence in a single class while assigning approximately equal confidence scores to two classes for hybrid cases.

\vspace{10pt}

\noindent To transform the predicted probabilities into an anomaly score, we first examine whether the maximum probability $p_{max_1}$ exceeds a predefined threshold $t$. If $p_{max_1} > t$, the butterfly is likely a non-hybrid from a single subspecies, and we define its anomaly score as $1 - p_{max_1}$. Conversely, if $p_{max_1} \leq t$, the specimen is considered a hybrid. In this case, we compute the anomaly score as the sum of the two highest probabilities, $p_{max_1} + p_{max_2}$, where $p_{max_2}$ is the second-highest probability. Formally, the anomaly score $A$ is defined as:

\begin{equation}
A=
\begin{cases}
    1 - p_{max_1}, & \text{if } p_{max_1} > t, \\
    p_{max_1} + p_{max_2}, & \text{otherwise}
\end{cases}.
\end{equation}


\begin{figure}
    \centering
    \begin{subfigure}[b]{0.3\linewidth}
        \centering
        \includegraphics[width=\textwidth]{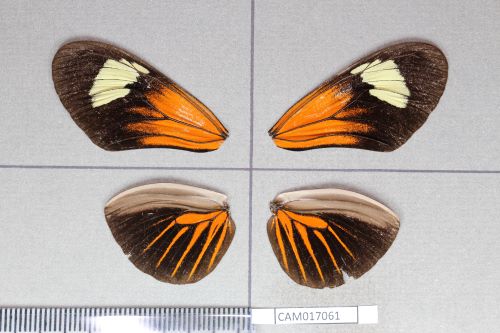}
        \caption{Subspecies I}
    \end{subfigure}
    \hfill
    \begin{subfigure}[b]{0.3\linewidth}
        \centering
        \includegraphics[width=\textwidth]{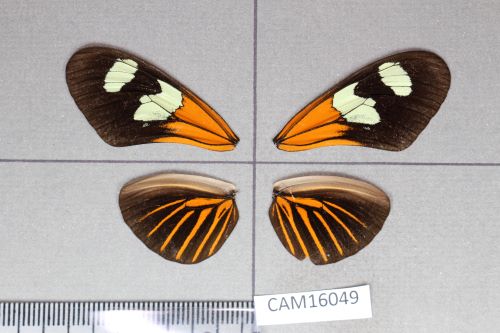}
        \caption{Hybrid cases}
    \end{subfigure}
    \hfill
    \begin{subfigure}[b]{0.3\linewidth}
        \centering
        \includegraphics[width=\textwidth]{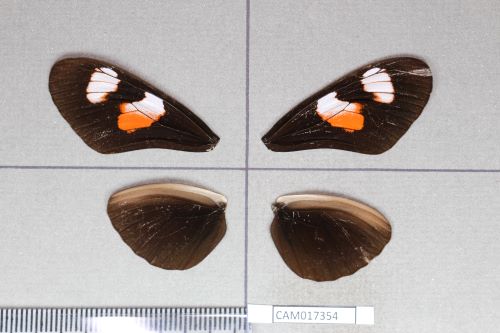}
        \caption{Subspecies II}
    \end{subfigure}
    \caption{Illustration of the hybrid and non-hybrid cases.}
    \label{fig:dataset}
\end{figure}

\begin{table}[t]
\centering
\caption{Results on the private dataset from the official evaluation platform of the top 6 team. $A$ denotes the AUC, and $R$ denotes the recall score. The subscript indicates whether the result corresponds to species A or B, while the superscript denotes hybrid and non-hybrid cases, respectively.}
\resizebox{\linewidth}{!}{
\begin{tabular}{lcccccccccc}
    \toprule
    \textbf{Participant} & $A_{A}^{H}$ & $A_{A}^{N}$ & \textbf{$R_B$} & $A^H_B$ & \textbf{Score $\uparrow$}\\
    \midrule
    anomalyfactory & 0.95 & 0.15  & 0.50 & 0.70 & 0.44\\
    Tater TOT & 0.99 & 0.13 & 0.76 & 0.89 & 0.37\\
    20iterations & 0.98 & 0.12 & 0.27 & 0.55 & 0.29\\
    kylechen1211 & 0.86 & 0.09 & 0.34 & 0.53 & 0.14\\
    motoki & 0.76 & 0.09 & 0.18 & 0.35 & 0.12\\
    \noalign{\smallskip} 
    \hdashline 
    \noalign{\smallskip}
    Ours & 0.99 & 0.21 & 0.68 & 0.81 & 0.44 \\
    \bottomrule
\end{tabular}
}
\label{tab:results}
\end{table}

\section{Experiment}

\subsection{Dataset} 

For training, we utilize the dataset provided by Imageomics for the NSF HDR Challenge\footnote{\url{https://www.codabench.org/competitions/3764}}. The dataset consists of 91 hybrid and 2001 non-hybrid images for species A. An illustration of the dataset is provided in Figure~\ref{fig:dataset}\footnote{Left to right: \texttt{CAM017061}, \texttt{CAM016049}, and \texttt{CAM017354}}. Model evaluation is conducted on the official platform using a private dataset including both species A and species B. Performance is assessed using recall and AUC as evaluation metrics.

\subsection{Setup} 

We train our model for 200 epochs with a batch size of 32 and set $t$ to 0.75. To prevent excessive drift in the backbone used for feature extraction, we set a low learning rate of $3 \times 10^{-6}$. Meanwhile, the classifier head is trained with a larger learning rate of $3 \times 10^{-4}$ to ensure effective adaptation. Since the mimic species exhibit similar coloration, we apply \textbf{color jittering} to enhance generalization. Additionally, to address the class imbalance between hybrid and non-hybrid samples, we employ a \textbf{weighted sampler} during training. As the provided images contain a significant amount of unused information, we further improve model robustness by applying \textbf{cropping} and \textbf{linear affine transformations} to refine feature extraction and reduce redundancy.

\subsection{Results}

We present the comparison results in Table~\ref{tab:results}. The results indicate that for species A, our method achieves the highest AUC for hybrid and non-hybrid classification. Regarding the mimicry performance, our model ranks second in both recall and AUC for species B hybrid cases. In terms of the overall score, we secure second place.

\begin{table}[t]
\centering
\caption{Ablation study of each component. The notations are the same to Table~\ref{tab:results}.}
\resizebox{\linewidth}{!}{
\begin{tabular}{lcccccccc}
    \toprule
    \textbf{Approach} & $A_{A}^{H}$ & $A_{A}^{N}$ & \textbf{$R_B$} & $A^H_B$ & \textbf{Score $\uparrow$}\\
    \midrule
    w. DINO+SVM  & 0.99 & 0.08 & 0.71 & 0.81 & 0.15 \\
    w. BioCLIP+SVM & 1.00 & 0.15 & 0.54 & 0.77 & 0.26 \\
    w.o jittering & 0.99 & 0.15 & 0.66 & 0.74 & 0.27 \\
    \noalign{\smallskip} 
    \hdashline 
    \noalign{\smallskip}
    Ours & 0.99 & 0.21 & 0.68 & 0.81 & \textbf{0.44} \\
    \bottomrule
\end{tabular}
}
\label{tab:ablation}
\end{table}

\subsection{Ablation Study}

To gain deeper insights into our approach, we conduct ablation study by evaluating each component also with the official evaluation platform as shown in Table~\ref{tab:ablation}.

\paragraph{Backbone.} We examine the backbone by comparing BioCLIP with DINOv2~\cite{oquab2024dinov}. The results demonstrate that BioCLIP achieves superior performance due to its strong capability to extract biologically relevant features. 

\paragraph{Probability Filtering.} We compare our probability filtering with an SVM-based approach~\cite{hearst1998support}. In this experiment, we freeze BioCLIP and train only the SVM for prediction. The results indicate that probability matching provides a more effective strategy for detecting hybrid cases.

\paragraph{Color Jittering.} We analyze the effect of color jittering and find that it further enhances performance by simulating variations observed in different mimicry species.

\section{Conclusion}

In this study, we developed an anomaly detection framework to identify hybrid butterflies and evaluated its generalizability to mimicry species. Our results demonstrate the effectiveness of BioCLIP for feature extraction and highlight the benefits of probability filtering. Notably, our approach achieved first place on the private dataset. Additionally, we conducted an ablation study to analyze the impact of each component on model performance, providing a comprehensive understanding of the proposed approach.

\bibliography{cite}

\begin{thebibliography}{3}
\providecommand{\natexlab}[1]{#1}

\bibitem[{Hearst et~al.(1998)Hearst, Dumais, Osuna, Platt, and Scholkopf}]{hearst1998support}
Hearst, M.~A.; Dumais, S.~T.; Osuna, E.; Platt, J.; and Scholkopf, B. 1998.
\newblock Support vector machines.
\newblock \emph{IEEE Intelligent Systems and their Applications}, 13(4): 18--28.

\bibitem[{Oquab et~al.(2024)Oquab, Darcet, Moutakanni, Vo, Szafraniec, Khalidov, Fernandez, HAZIZA, Massa, El-Nouby, Assran, Ballas et~al.}]{oquab2024dinov}
Oquab, M.; Darcet, T.; Moutakanni, T.; Vo, H.~V.; Szafraniec, M.; Khalidov, V.; Fernandez, P.; HAZIZA, D.; Massa, F.; El-Nouby, A.; Assran, M.; Ballas, N.; et~al. 2024.
\newblock {DINO}v2: Learning Robust Visual Features without Supervision.
\newblock \emph{Transactions on Machine Learning Research}.

\bibitem[{Stevens et~al.(2024)Stevens, Wu, Thompson, Campolongo, Song, Carlyn, Dong, Dahdul, Stewart, Berger-Wolf et~al.}]{stevens2024bioclip}
Stevens, S.; Wu, J.; Thompson, M.~J.; Campolongo, E.~G.; Song, C.~H.; Carlyn, D.~E.; Dong, L.; Dahdul, W.~M.; Stewart, C.; Berger-Wolf, T.; et~al. 2024.
\newblock Bioclip: A vision foundation model for the tree of life.
\newblock In \emph{Computer Vision and Pattern Recognition}, 19412--19424.

\end{thebibliography}

\end{document}